
\def\chaphead{}
\def\ni{\noindent}

\font\tfont=cmbxti10
\font\eightrm=cmr8
\font\eightit=cmti8
\font\sixrm=cmr6
\font\eightmit=cmmi8
\font\sixmit=cmmi6
\def\absmath{\textfont0=\eightrm \scriptfont0=\sixrm
	      \textfont1=\eightmit \scriptfont1=\sixmit}
\def\absfont{\let\rm=\eightrm \let\it=\eightit \rm\absmath}
\font\twelverm=cmr12
\font\twelveit=cmti12
\font\tenrm=cmr10
\font\twelvemit=cmmi12
\font\tenmit=cmmi10
\def\regmath{\textfont0=\twelverm \scriptfont0=\tenrm
	      \textfont1=\twelvemit \scriptfont1=\tenmit}
\def\peterfont{\let\rm=\twelverm \let\it=\twelveit \rm\regmath}
%
%

\newfam\vecfam

\textfont\vecfam=\tfont \scriptfont\vecfam=\seveni
\scriptscriptfont\vecfam=\fivei


\def\spose#1{\hbox to 0pt{#1\hss}}

\font\eightrm=cmr8

\def\s{\ifmmode \widetilde \else \~\fi} 
     
\def\section{\S}
\newcount\notenumber
\notenumber=1
\newcount\eqnumber
\eqnumber=1
\newcount\fignumber
\fignumber=1
\newbox\abstr


\def\s{{\rm\,s}}

\def\note#1{\footnote{$^{\the\notenumber}$}{#1}\global\advance\notenumber by 1}
\def\foot#1{\raise3pt\hbox{\eightrm \the\notenumber}
     \hfil\par\vskip3pt\hrule\vskip6pt
     \noindent\raise3pt\hbox{\eightrm \the\notenumber}
     #1\par\vskip6pt\hrule\vskip3pt\noindent\global\advance\notenumber by 1}

\def\abstract#1{\setbox\abstr=\vbox{\hsize 5.0truein{\par\noindent#1}}
    \centerline{ABSTRACT} \vskip12pt \hbox to \hsize{\hfill\box\abstr\hfill}}
     
\def\Dt{\spose{\raise 1.5ex\hbox{\hskip3pt$\mathchar"201$}}}    
\def\dt{\spose{\raise 1.0ex\hbox{\hskip2pt$\mathchar"201$}}}    

\def\new{{\rm\chaphead\the\eqnumber}\global\advance\eqnumber by 1}
\def\ref#1{\advance\eqnumber by -#1 \chaphead\the\eqnumber
     \advance\eqnumber by #1 }
\def\last{\advance\eqnumber by -1 {\rm\chaphead\the\eqnumber}\advance
     \eqnumber by 1}
\def\eqnam#1{\xdef#1{\chaphead\the\eqnumber}}
     
\def\nfig{\chaphead\the\fignumber\global\advance\fignumber by 1}
\def\nfiga#1{\chaphead\the\fignumber{#1}\global\advance\fignumber by 1}
\def\rfig#1{\advance\fignumber by -#1 \chaphead\the\fignumber
     \advance\fignumber by #1}
\def\fignam#1{\xdef#1{\chaphead\the\fignumber}}

\def\lta{\mathrel{\spose{\lower 3pt\hbox{$\mathchar"218$}}
     \raise 2.0pt\hbox{$\mathchar"13C$}}}
\def\gta{\mathrel{\spose{\lower 3pt\hbox{$\mathchar"218$}}
     \raise 2.0pt\hbox{$\mathchar"13E$}}}
     

\magnification=\magstep1
\parskip=3pt

\magnification=\magstep1
\def\ni{\noindent}
\def\refind{\noindent \hangindent=2pc \hangafter=1}

\centerline{\bf The structure of the central disk of NGC 1068: a clumpy disk 
model}
\bigskip
\centerline{\bf Pawan Kumar$^\dagger$}
\medskip
\centerline{Institute for Advanced Study, Princeton, NJ 08540}

\vskip 1.0cm
\centerline{\bf Abstract}
\bigskip
\baselineskip=15pt

NGC 1068 is one of the best studied Seyfert II galaxies, for which 
the blackhole mass has been determined from the Doppler velocities of 
water maser. We show that the standard $\alpha$-disk model of NGC 1068
gives disk mass between the radii of 0.65 pc and 
1.1 pc (the region from which water maser emission is detected) to 
be about 7x10$^7$ M$_\odot$ (for $\alpha=0.1$), more than four times
the blackhole mass, and a Toomre Q-parameter for the disk is $\sim$0.001. 
This disk is therefore highly self-gravitating and is subject to 
large-amplitude density fluctuations. We conclude that the standard 
$\alpha$-viscosity description for the structure of the accretion 
disk is invalid for NGC 1068.

In this paper we develop a new model for the accretion disk.
The disk is considered to be composed of gravitationally bound clumps;
accretion in this clumped disk model arises because of gravitational
interaction of clumps with each other and the dynamical frictional drag 
exerted on clumps from the stars in the central region of the galaxy. 
The clumped disk model provides a self-consistent description
of the observations of NGC 1068. The computed temperature and density are
within the allowed parameter range for water maser emission, and the 
rotational velocity in the disk falls off as $r^{-0.35}$.

\medskip
\ni{\it Subject heading:} accretion disks -- galaxies: individual (NGC 1068)
\vskip 9.3truecm
\noindent $^\dagger$ Alfred P. Sloan Fellow

\vfill\eject

\centerline{\bf 1. Introduction}
\bigskip

The discovery of emission from water masers in the central region of 
Seyfert II galaxy NGC 1068 provides an opportunity to study the 
properties of the associated accretion disk. The mass of the central 
black hole is estimated to be 1.5x10$^7$ M$_\odot$ (Greenhill \& 
Gwinn, 1997) from the measurement of the Doppler velocities of the 
masing spots.

The nucleus of NGC 1068 is heavily obscured, and the luminosity of the central
source is determined from the observed flux by modeling the dust obscuration
and the scattering of photons by ionized gas in the nuclear region. According
to a careful analysis carried out by Pier et al. (1994) the bolometric
luminosity of NGC 1068 is estimated to be about 8x10$^{44}$ erg s$^{-1}$;
the luminosity is perhaps uncertain by a factor of a few.

We use the blackhole mass and the bolometric luminosity to construct
the standard viscous disk model for NGC 1068 (\S2) and find that the
disk is highly self-gravitating thereby rendering the $\alpha$-disk model
inapplicable. The effect of the irradiation of the disk from the 
central source does not modify this conclusion (\S2b).

In section 3 we present a new model for the disk
in NGC 1068 composed of gas clumps. The accretion in this 
case arises as a result of gravitational interaction amongst clumps
(\S3.1); the structure of the clumped-disk model (CD model) of NGC 1068
is described in \S3.2.

The velocity of the masing spots appears to be falling off with distance from 
the center as r$^{-0.35}$ (Greenhill \& Gwinn, 1997), which is less rapid
than the Keplerian power law of $-0.5$. One possible reason for this could 
be that the disk of NGC 1068 is sufficiently massive
so as to modify the rotation curve. However, we show in \S3 that this is 
not so. Another possibility is that the flattening of the rotation curve
is due to a central star cluster. The influence of the star cluster
on the accretion rate and the disk structure is discussed in section 3.2.

\bigskip
\centerline{\bf 2. Standard thin disk model for NGC 1068}
\medskip

The standard viscous accretion disk model for NGC 1068, including
the irradiation of the disk from the central source, is presented below.
Throughout this paper we take the mass of the blackhole at the center of
NGC 1068 to be 1.5x10$^7$ M$_\odot$ (Greenhill \& Gwinn, 1997),
and the bolometric luminosity to be 8x10$^{44}$ erg s$^{-1}$ (Pier et al.,
1994).

The theory of thin accretion disk is well developed and is described in
a number of review articles and monographs e.g. Frank, King \& Raine (1992).
When the flux intercepted by disk is not small compared to the local
energy generation rate, such as that expected for the masing disk of
NGC 1068, the incident flux must be included in determining the thermal 
structure of the disk.

The fractional luminosity intercepted by the disk depends on the 
disk geometry, the scattering of radiation by the coronal 
gas etc. For instance if the dominant 
source of radiation intercepted by the disk were the scattered radiation from
an extended corona then we might expect the incident flux at the disk to be
roughly uniform. On the other hand for a flaring or a warped disk the radiation
intercepted from the source directly might dominate, and the incident 
radiation in this case depends on the inclination angle of the normal to
the disk. We adopt the second model to analyze the disk structure;
it is straightforward to consider other possibilities, 
but we do not pursue these since the main result of this section
viz., the standard $\alpha$-disk model for NGC 1068 is highly unstable to
gravitational perturbations turns out to be independent of whether we include
irradiation and radiative forces in the calculation of disk structure.

Let the luminosity of the central source be $L_c$. The flux incident at the
disk, $F_{in}$, a distance $r$ from the central source is taken to be

$$ F_{in}(r) = {L_c\over 4\pi r^2} {d H\over dr} = {\beta H 
   L_c\over 4\pi r^3}, \eqno(\new)$$

\ni where $H$ is the vertical scale height, the constant factor $\beta$ is 
defined by $dH/dr = \beta H/r$,
and determines the fraction of the flux that is intercepted by the disk. 
The upward energy flux due to mass accretion rate $\dot M$ is

$$ F_{up}(r) = {3\over 8\pi} \Omega^2 \dot M, \eqno(\new)$$
and the ratio of these fluxes $F_{in}/F_{up} \approx \epsilon\beta H/R_{sc}$,
where $\epsilon\approx 0.1$ is the efficiency of the conversion of the 
rest mass to energy by the central source, and $R_{sc}$ is the Schwarzschild 
radius of the blackhole, $\Omega=(GM_t/r^3)^{1/2}$ is the angular
rotation speed, and $M_t$ is the mass contained inside the radius $r$.
The effective temperature of the disk at $r$ is given by

$$ \sigma T_{eff}^4(r) = F_{up} + F_{in} \equiv F_t. \eqno(\new)$$

The temperature at the mid-plane of the disk ($T_c$) can be calculated by 
considering the first moment of the radiative transfer equation, and
making use of the Eddington approximation, and is given by

$$ T_c^4 = {3\over 4} T_{eff}^4 \left( {F_{up}\over F_t}\right) \left[
   \tau_0 + {4\over 3} {F_t\over F_{up}} - {2\over 3}\right], \eqno(\new)$$
where $\tau_0=\kappa\Sigma$ is the optical depth of the disk. The opacity
$\kappa$, for masing disks, with $H_2$ density and temperature in the 
range 10$^8$--10$^{10}$  cm$^{-3}$ and 100--1000 K respectively is
dominated by metal grains and is given by (Bell \& Lin 1994)

$$ \kappa \approx 0.1 \,T^{1/2} {\rm cm}^2 {\rm g}^{-1}. \eqno(\new)$$

The temperature in the disk mid-plane is not sensitive to the height where 
the energy is deposited so long as the optical depth of the disk to the 
incident radiation is much greater than one.

The solution to the hydrostatic equilibrium equation in the vertical 
direction, which includes the radiation pressure at the disk surface
$\sim (2F_{in}+F_{up})/c$, yields

$$ \rho_c c_s^2 \approx \Omega^2\Sigma H + {2F_t\over c}\left(1 -
   {\tau_0 F_{up}\over 2 F_t}\right). \eqno(\new)$$
where $\Sigma$ is the surface mass density, $c_s$ \& $\rho_c$ are the
sound speed and gas density in the mid plane of the disk.
Taking the effective viscosity in the disk to be $\alpha c_s H$ the 
mass accretion rate $\dot M$ is given by

$$ \dot M = {2\pi\alpha\Sigma c_s^2\over \Omega}. \eqno(\new)$$

Equations (\ref4), (\ref2) and (\ref1) are three equations in three
unknowns viz. $\Sigma$, $H$ and $T_c$, which are solved numerically and
the results are shown in fig. 1. Also shown in fig. 1, for 
comparison, are the solutions for $\beta=0$ i.e. when the irradience 
of the disk is neglected. The solution in the latter case can be obtained 
analytically and is given below

$$ T_c = 200\;\alpha^{-2/9} \dot m^{4/9} M_7^{7/9} r^{-1}\quad {\rm K}, 
   \eqno(\new)$$

$$ n = 10^{10} \alpha^{-2/3} \dot m^{1/3} M_7^{5/6} r^{-3/2}\quad {\rm cm}^{-3},
   \eqno(\new)$$
and the Toomre-Q parameter for the stability of the disk is

$$ Q = {\Omega c_s\over \pi G \Sigma} = 2.4\times 10^{-3} \alpha^{2/3}
   \dot m^{-1/3} M_7^{1/6} r^{-3/2}, \eqno(\new)$$
\ni where $\dot m$ is the accretion rate in terms of the Eddington rate, 
$M_7 = M/(10^7 M_\odot)$, and $r$ is the radial distance from the center in
parsecs. For the NGC 1068 system $M_7\approx 1.5$ and $\dot m\approx 0.4$. 
The inner and the outer radii of the masing disk are at 0.65pc and 1.1pc
respectively.

For the numerical results shown in fig. 1 we took the value of $\beta$ 
such that the disk intercepts about 50\% of the flux from the central source.
In this case the disk temperature is dominated by the incident flux, 
and the temperature in the disk midplane is close 
to $T_{eff}\approx 511$K at $r=1$pc. The structure of the disk ($\Sigma$, 
$Q$, $H$, $T_c$) in this case is almost independent of the opacity of 
the gas and consequently it is unaffected by any uncertainty in $\kappa$.
The molecular mass of the masing disk, when irradiation is included, 
is $\sim 7.0$x10$^{7} M_\odot$ (for $\alpha=0.1$), and the $Q$ is
about 10$^{-3}$ (see fig. 1). For the irradiation dominated disk, the 
disk mass decreases with $\alpha$
as $\alpha^{-0.9}$, and the $Q$ increases as $\alpha^{0.9}$. Since
$Q$ should be greater than one for stability, we see that the masing 
disk of NGC 1068 is highly unstable to gravitational perturbation. A
decrease in the irradiation flux makes the disk more unstable.

The main conclusion of this section is that the standard $\alpha$-disk
model for the NGC 1068 masing disk is inconsistent. The disk 
according to this model is highly unstable, and should fragment into 
clumps. A consistent analysis in this case should include the effect of 
self-gravity, and clump interaction, which is described below.

\bigskip
\centerline{\bf 3. A model for clumpy self gravitating disks}
\medskip
A disk with small value for ($1-Q$) is likely to develop spiral structure 
which can transport angular momentum outward. However, when $Q\ll1$, as 
in the case of NGC 1068 (see \S2), the disk is likely to breakup into clumps, 
and the accretion rate is determined by the gravitational interaction among
these clumps. Accretion in self gravitating disks was considered by 
Paczynski (1978), Lin \& Pringle (1987), Shlosman \& Begelman (1987), 
Shlosman et al. (1990), and has been investigated more recently by Kumar 
(1998) in some detail for clumpy disks. In \S3.1 we derive the results 
that we need to construct model for clumpy disks (CDs), and its
application to NGC 1068 is discussed in \S3.2.

\medskip
\ni {\bf \S3.1 Velocity dispersion and accretion rate in a clumpy disk}
\medskip

Consider a disk consisting of clumps of size $l_c$, and take the mean 
separation between clumps to be $d_c$. The tidal radius of a cloud, 
the distance to which the gravity of the cloud dominates over the gravity
of the central mass, is $d_t\sim r (m_c/M_t)^{1/3}$; where $m_c\approx 
\sigma_c l_c^2$ is the cloud mass, $\sigma_c$ is the surface mass 
density of the cloud and $M_t$ is the total axisymmetrically distributed 
mass contained inside the radius $r$. 

The change to the displacement amplitude of the epicyclic
oscillation of a clump as a result of gravitational interaction with
another clump, with impact parameter $d\gta d_t$, can be shown to
be $\sim m_c r^3/(M_t d^2)$. Gravitational encounters with $d\lta d_t$ 
are almost adiabatic and these leave the epicyclic energy of clumps 
unchanged (two particles moving on circular
orbits merely interchange their trajectories in such an encounter).
Thus the dominant gravitational interactions for exciting epicyclic 
oscillation are those with impact parameter of $\sim d_t$, and the 
change to the epicyclic amplitude in such an encounter is $\delta r 
\sim m_c r^3/(M_t d_t^2)\sim d_t$, or

$$ \delta r \sim d_c^{2/3} r^{1/3} \left({M_r\over\pi M_t}\right)^{1/3}, 
   \eqno(\new)$$
where 

$$ M_r = {\pi r^2\sigma_c l_c^2\over d_c^2} \equiv \pi r^2 \bar\sigma(r), 
   \eqno(\new)$$
and $\bar\sigma(r)$ is the mean surface mass density of the disk at radius $r$.

The mean time interval for a clump to undergoes strong gravitational 
interaction with another clump i.e. with impact parameter $d_t$ is 
$t_{int} \sim \Omega^{-1} (d_c/d_t)^2$, and so long as the cloud size
($l_c$) is not much smaller than $d_c$ the timescale for physical
collision between clumps is also of the same order i.e.
clouds undergo physical collision after having undergone one strong 
gravitational encounter on the average, and so the cloud velocity dispersion 
is $\sim \Omega\delta r$ or

$$ \bar v_{r}\sim (r\Omega) \left[{M_{r}\over \pi M_{t}}\right]^{1/3}
   \left[{d_c\over r}\right]^{2/3}, \eqno(\new)$$
The expression for $\bar v_r$ is similar to that given by Gammie et 
al. (1991) for the velocity dispersion of molecular clouds in the Galaxy.

We assume that the kinetic energy of the relative motion of colliding 
clouds is dissipated and their orbit is circularized. 
Colliding clumps might coalesce so long as the size of the resulting
object does not exceed the maximum length
for gravitational instability, $l_{max}\sim(rM_r)/M_t$; clouds of larger
size are susceptible to fragmentation due to the rotational shear of the
disk which limits their size.

The characteristic time for a clump to fall to the center can be estimated
using equation (\ref3) and is given by

$$ t_r \approx \Omega^{-1} \left({d_c\over r}\right)^2 \left({M_t\over m_c}
   \right)^{4/3}, \eqno(\new)$$
\ni and the associated average mass accretion rate is

$$ \dot M \approx \Omega(r) M_r \left({M_r\over \pi M_{t}}\right)^{4/3}
   \,\left({d_c\over r}\right)^{2/3}. \eqno(\new)$$
This result is applicable so long as the density contrast
between the clump and the inter-clump medium is about a factor of two
or larger, and $l_c$ is of order $d_c$.

For $l_c\ll d_c$ clouds undergo several gravitational encounters
with other clouds before undergoing a physical collision, and during 
this time the random velocity of clouds continues to increase. 
The rate of increase of velocity dispersion in two body gravitational
interaction is given by

$$ {d v^2\over dt} \approx {G^2 m_c M_r \Omega\over r^2 v^2}. \eqno(\new)$$
We assume that the scattering gives rise to isotropic velocity dispersion,
and so the scale height for the vertical distribution of clumps is
$\sim v/\Omega$. The time for a clump to undergo physical collision, 
assumed to be completely inelastic, is $t_{col}\sim \Omega^{-1} (d_c/l_c)^2$, 
and therefore the velocity dispersion is  

$$ \bar v^2 \approx {G m_c\over l_c}. \eqno(\new)$$
This corresponds to the Safronov number being equal to one.
Since the cloud collision time is much greater than $\Omega^{-1}$ 
the effective viscosity is suppressed by a factor $(\Omega t_{col})^2$ compared
to the case where collision frequency is greater than $\Omega$,
(cf. Goldreich and Tremaine, 1978), i.e. $\nu_e \sim \bar v_r^2 
(t_{col}^{-1}/\Omega^2)$, and thus the mass accretion rate is given by

$$ \dot M \approx M_r\Omega \left({M_r\over M_t}\right) \left({l_c\over r}
   \right)\approx \Omega M_r \left({M_r\over M_{t}}\right)^{4/3}
   \,\left({d_c\over r}\right)^{2/3} \left({l_c\over d_t}\right). \eqno(\new)$$
This accretion rate is smaller than given by equation (\ref4) by a factor
of $(d_t/l_c)$.

We can parameterize the effect of unknown size distribution and separation 
between clumps on the accretion rate and the velocity dispersion by a 
dimensionless parameter $\eta$, and rewrite equations for $\dot M$ 
and $\bar v_r$ in the following form for future use:

$$ \dot M = \eta \Omega(r) M_r(r) \left({M_r\over \pi M_{t}}\right)^2,
    \eqno(\new)$$
and 
$$ \bar v_r = \eta r\Omega(r) \left({M_r\over \pi M_t}\right). \eqno(\new)$$

For $l_c\sim d_c\sim l_{max}$, we see from equations (16) and (18) that
$\eta\sim 1$. The effective kinematic viscosity in this case is 
approximately $\l_{max}^2\Omega\sim Q^{-2} H^2 \Omega$ (where $Q\sim 
l_{max}/H$ is the Toomre $Q$-parameter, $H\sim \bar v_r/\Omega$), same as in
the ansatz suggested by Lin \& Pringle (1987) for self-gravitating-disk. 
We note that the relative velocity
of collision between clouds is smaller than their orbital speed by a factor
of the ratio of the total mass to the molecular mass, and as long as the 
cooling time for post-collision gas is less than the time between collisions
the clouds are not smeared away due to heating and the disk remains clumpy.

\bigskip
\ni{\bf \S 3.2 Application to NGC 1068}
\medskip

The equation (\ref2) can be recast in the following form

$$ {M_r^2\over M_r + M_c} = \psi, \eqno(\new)$$
where
$$ \psi \equiv {\pi^2\over\eta^{2/3}} {r \dot M^{2/3}\over G^{1/3}}.
   \eqno(\new)$$
The solution to this equation in the case where the central mass 
dominates is 

$$ M_r \approx \psi^{1/2} \left( M_c^{1/2} + {\psi^{1/2}\over 2}\right),
   \eqno(\new)$$
and from equation (\ref4) we find the velocity dispersion of clouds to be

$$ \bar v_r \approx \eta^{2/3} [G \dot M]^{1/3}. \eqno(\new)$$

For NGC 1068, the luminosity $\sim$ 8x10$^{44}$ erg s$^{-1}$, $M_c\sim$
1.5x10$^7$ M$_\odot$, and $\dot M\sim 8$x10$^{24}$ g s$^{-1}$.
Substituting these numbers in the above equations we find

$$ M_r \approx 1.3\times 10^6 \; M_\odot\; r^{1/2} \eta^{-1/3}, \eqno(\new)$$
and
$$ \bar v_r \approx 8 \, {\rm km}\, {\rm s}^{-1} \; \eta^{2/3}, \eqno(\new)$$
where $r$ is measured in parsecs. Note that the disk mass in this model is more
than an order of magnitude smaller than in the standard $\alpha$-disk model
discussed in \S2. The velocity dispersion of clumps is larger than the sound
speed, and is of order the observed scatter in the velocity
of masing spots in NGC 1068.

The number density of molecules and the gas temperature can be obtained by
solving the hydrostatic and the thermal equilibrium equations for blobs 
in the vertical direction in the presence of incident flux (see \S2).
We find the thermal temperature of blobs to be about 510 K at $r=1$pc
(see fig. 1), and the density scale height $H\sim 1.5\times10^{16}$ cm.
The mean number density of $H_2$ molecules $n\sim M_r/(4\pi r^2 H \mu)= 
5\times 10^8$ cm$^{-3}$. Thus both the number 
density and the temperature are within the allowed range for water maser
emission.\footnote{$^1$}{The allowed range for number density for water
maser emission is 10$^8$---10$^{11}$ cm$^{-3}$ and the temperature range
is 200---1000 K.}
The size of clumps is $\lta r (M_r/M_t)\sim 0.1$ pc and their mass is
$\sim 10^3$ M$_\odot$. Since the Jeans mass is $\sim 10$ M$_\odot$ the
clumps could have some star formation activity. However, the efficiency
of star formation is usually quite low, of order a few percent even for
the giant molecular clouds, so we don't expect the gas in the clumps
to be converted into stars in the short lifetime of order 10$^6$ years
or less for the clumps in the masing disk of NGC 1068.

Note that the velocity dispersion of blobs is independent of $r$ (eq.
[\ref1]), so the scale height for the vertical distribution of blobs
increases as $r^{3/2}$ provided that the velocity distribution of clouds
is nearly isotropic; this increase of disk thickness with $r$ 
is more rapid than in the standard $\alpha$-disk model considered in the
last section. At a distance of 2 pc from the center of NGC 1068 the scale 
height is about 0.3 pc (for $\eta=5$), so the flaring disk blocks a 
significant fraction of the radiation from the central source.

Clouds colliding at a relative speed of 8 km s$^{-1}$ raise the gas 
temperature to approximately 10$^4$ K which is too hot for maser emission.
However, the cooling time of gas at the density of $\sim 10^9$ cm$^{-3}$, is
of order 100 years, which is short compared to both the orbital time and the
collision time of order $10^4$ years\footnote{$^2$}{The density enhancement of
the shocked gas depends on the strength of magnetic field in the clumps and is
order unity for equipartition magnetic field.} ensuring that the disk
remains cold and that a steady state solution exists.

The slope of the rotation curve in the disk follows from the use of
equation (\ref2)

$$ {d\ln V_{rot}\over d\ln r} = -0.5 + {1\over 2}{d\ln M_t\over d\ln r} = -0.5
    + {\psi^{1/2}\over 4 M_c^{1/2}} \approx -0.5 + {M_r\over 4 M_c}.
   \eqno(\new)$$
The disk mass for the NGC 1068 system, in our model, is small 
($M_r/M_t\approx 0.1$), and so the slope of the rotation curve is 
very close to $-0.5$.

A slower fall off of the rotation curve requires a more rapid increase of
the total mass ($M_t$) with $r$ than in the model discussed above.
This could arise for instance if there is a star cluster at the center.
The rotation curve falls off as $r^{-0.35}$ when the mass of the star cluster 
within a parsec of the center of NGC 1068 is about 8x10$^6$ M$_\odot$.
Thatte et al. (1997) have carried out near infrared speckle imaging of the
central 1" of NGC 1068, and conclude that about 6\% of the near-infrared
light from this region is contributed by a star cluster. They
estimate that the stellar mass within 1" ($\sim 50$pc) of the nucleus
of NGC 1068 is about 6x10$^8$ M$_\odot$; if the density in this cluster 
falls off as $r^{-2}$, as in a singular isothermal sphere, then the 
expected stellar mass inside 1 pc is 1.2x10$^7$ M$_\odot$. We note that
the stellar mass within 100 pc of our own galactic center is estimated to
be in excess of 5x10$^8$ M$_\odot$ (Kormendy \& Richstone 1995, Genzel et al.
1994), and the mass enclosed within radius $r$ is seen to increase almost 
linearly with $r$ for $r\gta 2$pc. Thus the possibility of a similar 
stellar mass cluster at the center of NGC 1068 is not surprising. 

We discuss below the effect a
star cluster has on the structure of the accretion disk.

\bigskip
\ni{\bf \S 3.2.1  Effect of a star cluster on accretion disk}
\medskip

Let us consider that the stellar mass within a radius $r$ of the center 
is $M_s(r)$. The disk mass, as before, is taken to be $M_r(r)$, and the central
mass is $M_c$. The use of equation (\ref5), which still applies with $M_c$ 
replaced by $(M_c+M_s)$, implies that the disk structure is not much affected
at small radii where $M_s(r)\ll M_c$. At larger radii, where the stellar
mass becomes comparable to or exceeds $M_c$, the mass of the gas disk
($M_r$) increases with $r$ as $r^{1/2} (M_c+M_s)^{1/2}$, which is somewhat
more rapid than the case of $M_s=0$ considered in \S3.1. 
However, for the masing disk of NGC 1068 $M_s\lta M_c$ for 
$r\lta 1.6$ pc, and the effect of the stellar cluster on 
the disk structure, in particular the number density of molecules and the gas
temperature, is small. 

The velocity dispersion of clouds due to gravitational encounters is
proportional to the local surface mass density $\bar\sigma$ of gas
and is unaffected by the stellar cluster
(see eq. [\ref4]). Therefore, the disk thickness (H)
at first increases with radius as $r^{3/2}$ and then flattens out when 
$M_s$ starts to dominate the total mass:

$$ H \approx {\bar v_r\over \Omega(r)} \approx {\eta r^3 \bar\sigma(r)\over
    M_t}. \eqno(\new)$$

\ni For $\bar\sigma\propto r^{-3/2}$, expected of the CD model with constant
$\dot M$, $H/r\propto r^{1/2}$, however a more rapid decrease of $\bar\sigma$
with $r$ leads to a corresponding decrease of $H/r$.\footnote{$^3$}{
A decrease of $\bar\sigma$ with $r$ that is more rapid than $r^{-3/2}$
leads to a drop in $\dot M$. However, the assumption of steady state 
accretion breaks down beyond some radius where
the accretion time, $\Omega^{-1} (M_c+M_s)^2/M_r^2$, is longer compared
with the evolution time scale of the disk.}

The accretion rate calculated in \S3.1 is modified due to the dynamical
friction suffered by clumps in the disk by the star cluster; the orbits
of stars in the cluster are gravitationally perturbed by the clumps
so that on the average the density of stars behind a clump is greater
than the density in the front. We estimate below the accretion rate
that arises as a result of the frictional drag exerted by stars. We assume
that clumps are not stretched out by the tidal force of stars, which is
valid so long as the cloud mass density is greater than the mean mass density 
associated with the star cluster i.e. $n_{H_2} \gta M_s/(m_{H_2} r^3)$
or $n_{H_2} \gta 2\times 10^8 \,{\rm cm}^{-3}\; (M_s/10^7 M_\odot)
(r/1 pc)^{-3}$; this condition is satisfied for NGC 1068.

The dynamical friction timescale for a clump to fall to the center
is (cf. Binney and Tremaine, 1987)

$$ t_{df} \approx {r^2 M_t v_{rot}\over G m_c M_s\ln\Lambda}, \eqno(\new)$$
where $m_c$ is clump mass, $\Lambda\approx r v_{rot}^2/(G m_c)$, and
$v_{rot}^2=GM_t/r$. Taking the clump size to be the largest length scale for
gravitational instability in a shearing disk i.e. $l_c\approx \pi^2 G
\bar\sigma/\Omega^2$, we find $m_c\approx \pi M_r^3/M_t^2$. Thus, $\ln\Lambda
\approx 3\ln(M_t/M_r)$ is about 7 for the NGC 1068 system.
Substituting these in the above equation we find

$$ t_{df} \approx {\Omega^{-1} \over 10} \left( {M_t^4\over M_s M_r^3}\right).
   \eqno(\new)$$
For $M_s\approx M_t/3$ and $M_r/M_t\approx 0.1$, values applicable to the
NGC 1068 system, the dynamical friction time is of the same order as the
timescale for clumps to fall to the center due to gravitational encounter
with other clumps (see eq. [22]). Thus we see that a star cluster of modest
mass at the center of NGC 1068 can both give rise to a sub-Keplerian rotation
curve, as perhaps observed by the water maser, and also because
of its dynamical friction on clumps remove angular momentum of the 
gaseous disk resulting in accretion
rate that is of the same order as needed to account for the observed
luminosity (the total $\dot M$ is the sum of the accretion rate due
to gravitational interaction between clumps, and the dynamical
friction drag exerted by stars).

It was pointed out by Ostriker (1983) that even a smooth disk is subject
to frictional drag from a star cluster. He showed that the characteristic
drag time for removing angular momentum of a gaseous disk is of the same
order as the relaxation time for the star cluster. The accretion rate due
to this process, in a form applicable to the NGC 1068 system, is given by 

$$ \dot M\sim M_r \Omega I_0 \left( {r_*\over r}\right)^2 \left( {M_s\over m_*}
   \right),   \eqno(\new)$$
where $r_*$ is the radius and $m_*$ is the mass of stars in the cluster, and
$I_0\sim 10$ is a dimensionless quantity that depends on the ratio of the
escape velocity at the surface of a star to the stellar velocity dispersion.
We see that the accretion rate resulting from
friction drag from stars on a smooth disk is much smaller than the rate 
resulting from the frictional drag on clumps calculated earlier.

\bigskip
\centerline{\bf 4. Conclusion}
\medskip
We find that the structure of masing disk of NGC 1068, as determined using 
the standard $\alpha$-viscosity prescription, is both inconsistent 
with the observations and is self contradictory. In particular, the disk
mass according to this model is 7x10$^7$ M$_\odot$, which is much greater
than the blackhole mass, and the Toomre $Q$ parameter has an extremely 
low value of $\sim 10^{-3}$ (for $\alpha=1$). This makes the disk highly 
unstable to gravitational perturbations, and suggests that it breaks up into 
clumps, a conclusion that is inconsistent with the smooth disk assumption of
standard $\alpha$-disks.

We have described in this paper a different model for the disk. 
We consider the disk to consist of 
gas clumps which undergo gravitational interactions with one
another leading to an inward accretion of gas. The mass of the
masing disk of NGC 1068 according to this model is about 
10$^6$ M$_\odot$, and the velocity dispersion of clumps is about 
10 km s$^{-1}$ which is in rough agreement with the observed 
velocity dispersion of masing spots. The temperature and the density
of clumps in this model are approximately 510 K and 5x10$^8$ cm$^{-3}$
respectively, which are within the allowed parameter range
for maser emission.

However, the rotational velocity of clumps, in this model, are close
to the Keplerian value, whereas the observations of masing spots indicate 
a slower fall off of their rotational velocity (Greenhill \& Gwinn, 1997). 
One obvious way these results can be reconciled is if the mass in stars
within 1 pc of the center of NGC 1068 is of order the blackhole
mass. Speckle observations of the central 1" region of NGC 1068 in near
infrared in fact suggests that the stellar mass contained within 50 pc of
the center is about 5x10$^8$ M$_\odot$ (Thatte et al. 1997), which is 
sufficient to explain the flattening of the observed rotation curve in the 
masing disk. The dynamical friction exerted by this star cluster on the 
clumps in the disk removes angular momentum at a rate that is of the same 
order as needed for the nearly Eddington accretion rate for the system.
Thus the clump-clump gravitational interaction and the dynamical
friction drag force on clumps together determine the disk
structure, which we find to be consistent with observations.

The clumpy-disk (CD) model considered in this paper for the 
disk of NGC 1068 should also apply to any AGN at a distance of about a pc
or more from the center, depending on the accretion rate and the luminosity, 
where the disk becomes self-gravitating and the standard
$\alpha$ model is too inefficient to account for the accretion rate
(Shlosman \& Begelman, 1989).

\medskip
\ni{\bf Acknowledgment:} I am indebted to Ramesh Narayan for
sharing his work and insights on the accretion disk
of NGC 1068, and for his detailed comments on this work.
I am grateful to Scott Tremaine for very helpful discussion about disk dynamics.
I thank John Bahcall, Mitch Begelman, Kathryn Johnston, Douglas Richstone, 
Phil Maloney and an anonymous referee for helpful comments.

\vfill\eject
\centerline{\bf REFERENCES}
\bigskip

\ni Bell, K.R. \& Lin, D.N.C., 1994, ApJ 427, 987

\refind Binney, J.J. \& Tremaine, S.D., 1987, {\it Galactic Dynamics}, Princeton 
University Press, Princeton

\refind Frank, J., King, I.R., \& Raine, D., 1992, Accretion Power in
 Astrophysics, (Cambridge: Cambridge univ. press)

\refind Gammie, C., Ostriker, J.P., and Jog, C., 1991, ApJ 378, 565

\ni Genzel, R., Hollenbach, D. and Townes, C.H., 1994, Rep. Prog. Phys. 57, 417

\ni Goldreich, P., and Tremaine, S., 1978, ICARUS 34, 227

\ni Greenhill, L.J., \& Gwinn, C.R., 1997, CfA preprint no. 4508

\ni Kormendy, J. \& Richstone, D., 1995, ARA \& A 33, 581

\ni Lin, D.N.C., and Pringle, J.E., 1987, MNRAS 225, 607

\ni Ostriker, J.P., 1983, ApJ 273, 99

\refind Paczynski, B., 1978, Acta Astr. 28, 91

\refind Pier, E.A., Antonucci, R., Hurt, T., Kriss, G., and Krolik, J., 1994,
ApJ 428, 124

\refind Shlosman, I., and Begelman, M., 1987, Nature 329, 810

\refind Shlosman, I., and Begelman, M., 1989, ApJ 341, 685

\refind Shlosman, I., Begelman, M., and Frank, J., 1990, Nature 345, 679

\refind Thatte, N., Quirrenbach, A., Genzel, R., Maiolino, R. \& Tecza, M.,
1997, ApJ 490, 238

\vfill\eject
\centerline{\bf Figure Caption}
\bigskip\ni
{\bf FIG. 1.---} The panel on the left shows the gas temperature ($T_c$) and
the number density of $H_2$ molecules ($n_c$) in the mid plane of the disk 
for the standard $\alpha$-disk model without any irradiation from the central 
source (thin continuous line, and thin dashed line respectively).
The thick continuous and dashed lines are $T_c$ and $n_c$ respectively
for an $\alpha$-disk model that includes irradiation from the central source. 
The irradiation is taken to be proportional to $H/r$, where $H$ is the scale 
height of the disk, and the proportionality constant ($\beta$) is chosen 
so that the flux intercepted by the disk at distance $r$ is about 50\% of 
the flux from the central source at $r$ (such a large value for the
flux intercepted by the disk corresponds to $\beta=50$ which might arise if
the disk were extremely warped). The panel on the right shows the 
Toomre Q-parameter for the two disk models; thin continuous curve is for the
standard $\alpha$-disk with no irradiation, and the thick curve is for
a disk that intercepts about 50\% of the flux. Note that $Q$ is much
less than one in both of these models which corresponds to the disk
being highly unstable to gravitational perturbations. In both of these 
models we chose $\alpha=0.1$ ($Q$ scales as roughly $\alpha^{0.9}$
when irradiation dominates),
the blackhole mass $M=1.5\times 10^7 M_\odot$, and the mass accretion 
rate ($\dot m$) in units of the Eddington rate was taken to be 0.4.

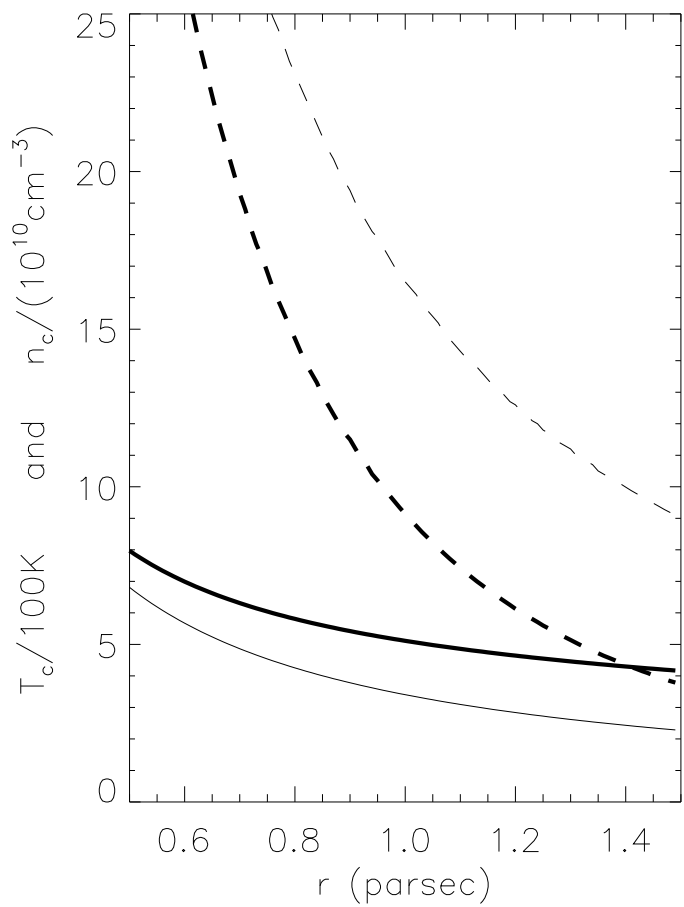
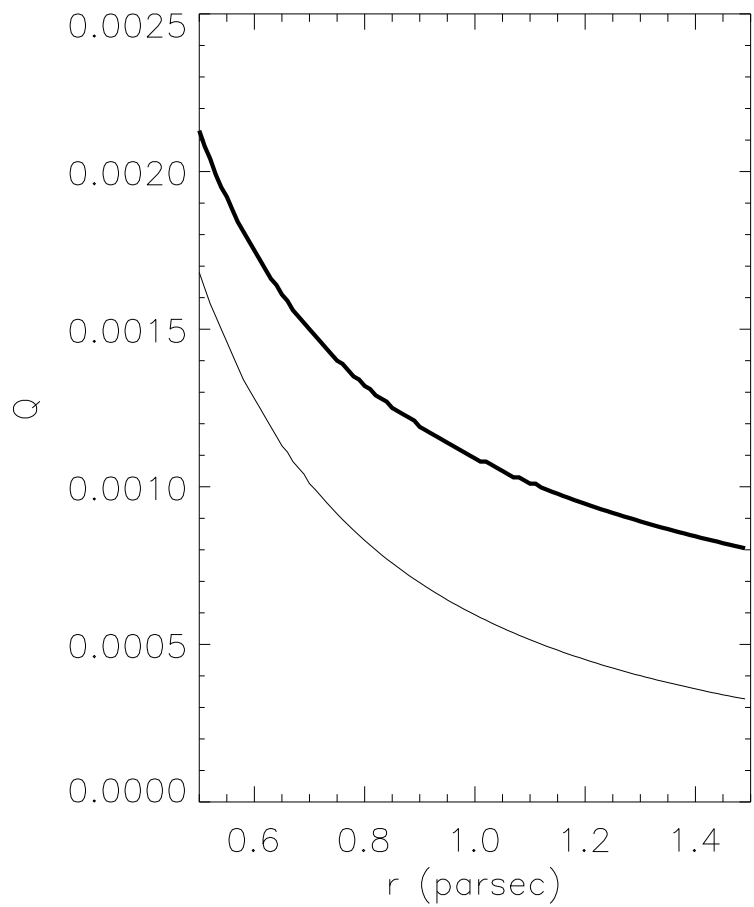
\end